\documentclass[10pt,twocolumn]{article}
\usepackage{sp_template}

\usepackage{braket}
\usepackage{dcolumn}
\usepackage{bm}
\usepackage{nicefrac}

\hypersetup{
  pdftitle  = {The phase boundary of the random site Ising model},
  pdfauthor = {R. Ben Ali Zinati, G. Gori, A. Codello},
}

\title{The phase boundary \\ of the random site Ising model}

\author{%
  R. Ben Al\`i Zinati\affilmark{1}%
  \and
  G. Gori\affilmark{2,3}%
  \and
  A. Codello\affilmark{4,5}%
}

\affiliation[1]{Universit\`a della Svizzera italiana, Via Giuseppe Buffi 13, Lugano 6900, Switzerland}
\affiliation[2]{CNR-INO, Area Science Park, Basovizza, 34149 Trieste, Italy}
\affiliation[3]{Institut f\"ur Theoretische Physik, Universit\"at Heidelberg, 69120 Heidelberg, Germany}
\affiliation[4]{DSMN, Ca'\ Foscari University of Venice, Via Torino 155, 30172 Venice, Italy}
\affiliation[5]{IFFI, Universidad de la Rep\'ublica, J.H.y Reissig 565, 11300 Montevideo, Uruguay}

\date{\today}
\begin{document}

\twocolumn[{%
  \maketitle
  \begin{abstract}
We introduce a new approach to disordered two-dimensional Ising models based on the extension of the Feynman--Vdovichenko combinatorial solution to randomized supercells.
Applying it to the site-diluted Ising model on the square lattice, we resolve the full phase boundary $T_c(p)$ from the pure-Ising point to the percolation limit $T_c(p_c)=0$ with, in principle, arbitrary precision.
The critical eigenvalue governing the transition is found to follow a remarkably accurate linear interpolation between the Ising and percolation endpoints, whose small but systematic deviations reveal the nontrivial fine structure of the phase boundary. Near the percolation threshold, we confirm the crossover exponent $\phi_{\rm RSIM}=1$ and extract the nonuniversal amplitude ${\alpha_{\rm RSIM}\simeq 1.616}$.
\end{abstract}
}]

%───────────────────────────────────────────────────────────────────────────────
\section*{Introduction}
The two-dimensional randomly site-diluted Ising model (RSIM) on the square lattice is a prototypical system of quenched disorder.
For site occupation probability $p$
above the percolation threshold $p_c=0.592746\ldots$ \cite{Newman2000,Jacobsen_2014}, the system undergoes a continuous paramagnetic--ferromagnetic transition, while for $p<p_c$ long-range order is absent.
The transition line $T_c(p)$ connects the pure Ising point $T_c(1) = 2/{\log(1+\sqrt{2})}$
to the percolation limit $T_c(p_c)=0$.
Along this line, the critical behavior belongs to the Ising universality class independent of dilution, with quenched disorder acting as a marginally irrelevant perturbation that produces universal logarithmic corrections to scaling \cite{Harris1974a,Dotsenko1983, Ludwig1987}.
Despite extensive numerical and analytical effort \cite{FishHarris1976,deSouza1992,Kim1995,Prudnikov1995,Ballesteros1997,Schreiber2005,Martins2007,Hasenbusch2008,Malakis2008,Yeomans1978,Yeomans1979,Jayaprakash1978,Tsallis1980,Plascak1991,Dotsenko1983,Ludwig1987,Plechko1997bj,dotsenko2017self,Rushbrooke1972,Queiroz1992,Kuhn1994,Mazzeo1999,Fulco2001,Kutlu2013}, the transition line has been determined only approximately or at a limited number of points, mostly far from the percolation threshold $p_c$; its complete determination remains open.
The only available exact results for site dilution are the initial derivative $T'_c(1)$ and successive terms of the expansion in $q=1-p$ \cite{AuYang1976,thorpe1976,thorpe1979phase}, and the asymptotic behavior near $p_c$ \cite{Coniglio1981}, including weak bounds on $T_c(p)$ \cite{Harris1974b, Bergstresser1976}.
For bond dilution the situation is somewhat more constrained, since $p_c=\tfrac{1}{2}$ is known exactly \cite{Sykes1963} and the asymptotic behavior is more stringent \cite{Harris1974a, Bergstresser1976,Domany1978}.
In this Letter we propose a new approach to disordered Ising models and use it to determine the full phase boundary $T_c(p)$ of the RSIM on the square lattice with, in principle, \emph{arbitrary precision}, establishing for the first time the complete connection between the pure-Ising and percolation limits.
\begin{figure}[th!]
    \centering
    \includegraphics[width=\columnwidth]
        {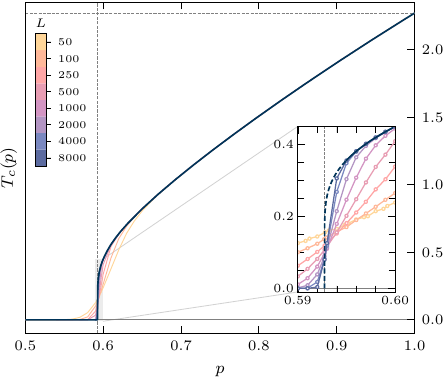}
    \caption{Phase boundary $T_c(p)$ of the site-diluted Ising model
  computed for increasing supercell size $L$. The inset magnifies the
  critical region near the percolation threshold $p_c$. The darker curve
  shows the converged  boundary: the asymptotic form
  $-2/\log[\alpha_{\mathrm{RSIM}}(p-p_c)]$ of Eq.~\eqref{asymp}
  near $p_c$ (dashed in the inset) with $\alpha_{\rm RSIM}\simeq 1.616$ and the converged data for
  $p \gtrsim 0.595$ (solid), where finite-$L$ curves collapse within
  numerical precision. Statistical errors are smaller than the symbol
  size.}
    \label{phasediagram}
\end{figure}
\begin{figure*}
    \centering
\includegraphics[width=1\textwidth]
{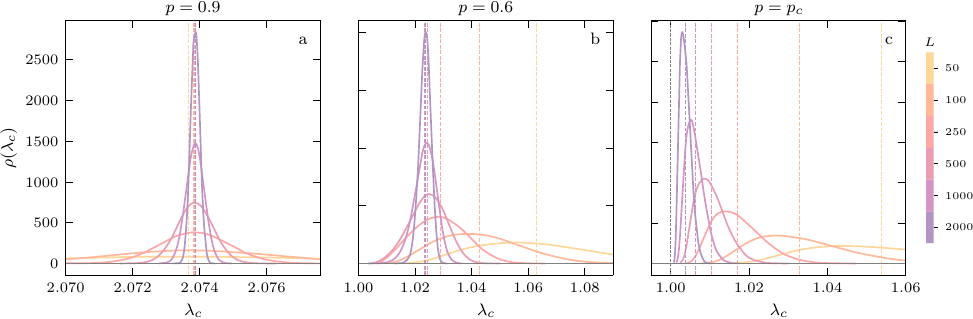}
    \caption{Distributions of the FV eigenvalue $\lambda_c$ for increasing supercell size $L$ at three representative dilutions: (a) $p=0.9$, (b) $p=0.6$, and (c) $p=p_c$.
    For high occupancy ($p=0.9$) the distributions rapidly converge, indicating complete self-averaging of $\lambda_c$, while at intermediate dilutions ($p=0.6$), convergence is slower and requires larger supercells.
    At the percolation threshold $p_c$ distributions remain size-dependent, indicating that convergence is not yet attained.
    }
    \label{convergence}
\end{figure*}

\vspace{5pt}
\section*{A new approach to disorder} Our approach builds on the Feynman--Vdovichenko (FV)  combinatorial solution of the two-dimensional Ising model \cite{Vodvicenko_1965, Feynman_1972, Kac_Ward_1952}.
In this formulation, the partition function on any periodic lattice $\Lambda$ with a finite unit cell is expressed in terms of a transition matrix $\mathbb{W}(\Lambda)$
whose spectrum encodes the critical condition \cite{Codello2010, Laurent_2025, Portillo_2025, Joseph_2026}.
The critical temperature $T_c$ is determined by the real eigenvalue $\lambda_c$ of $\mathbb{W}$ satisfying $1/\lambda_c \in (0,1)$
via $\lambda_c = 1/\tanh(1/T_c)$ (we set $k_B=J=1$).
This spectral construction extends naturally to lattices built from $L\times L$ supercells, as previously exploited to approximate the critical temperatures on non--homogeneous fractal lattices \cite{Codello:2015bia,Perreau2017,Zinati:2025npw}.

The key insight of this Letter is that this formulation can be adapted to quenched disorder:
sites within each supercell are randomly deactivated with probability $1-p$, defining a disordered transition matrix $\mathbb{W}_{L,p}^{(r)}$ for each disorder realization $r$.
The corresponding critical temperature $T_c^{(r)}(L,p)$ is then determined exactly by the spectral condition
\begin{equation}\label{eq:criticalcond}
    \det(\lambda_c^{(r)}(L,p)\,\mathbb{I} - \mathbb{W}_{L,p}^{(r)}) = 0\,.
\end{equation}
The disorder-averaged critical temperature
\begin{equation}
    \label{eq:avgT}
    \overline{T}_c(L,p) = \frac{1}{N} \sum_{r=1}^N T_c^{(r)}(L,p) \,,
\end{equation}
converges to the thermodynamic phase boundary $T_c(p)= \lim_{L\to \infty} \overline{T}_c(L,p)$ as the supercell size increases.
A similar relation defines $\overline{\lambda}_c(L,p)$ and $\lambda_c(p)$.
The sum in Eq.~\eqref{eq:avgT} extends over all $N$ realizations, including both percolating and non-percolating configurations; for the latter the spectral condition yields $T_c^{(r)}(L,p)=0$, so that $\overline{T}_c(L,p)$ consistently incorporates both dilution and percolation effects.
Statistical errors arise solely from the disorder average in Eq.~\eqref{eq:avgT}.
At each supercell size $L$ and concentration $p$, the number of samples $N$ is chosen such that the standard error of the mean falls below a prescribed tolerance, ensuring that finite-sampling fluctuations are negligible on the scale of our results.
The resulting phase boundary is shown in Fig.~\ref{phasediagram} for increasing supercell sizes, together with the limiting curve.

\vspace{5pt}
\section*{Derivative at the pure limit}
The derivative of the critical line at the pure Ising point is known exactly,
$T'_c(1)=
\tfrac{4\pi}{(\sqrt{2}+\pi)\log(1+\sqrt{2})^2}
=3.550787\dots$ \cite{thorpe1979phase,AuYang1976,Dotsenko1983}, providing a stringent benchmark for our approach.
Within our framework $T'_c(1)$ is obtained from the finite-difference estimator
\begin{equation}
T'_c(1)=\lim_{L\to\infty}
\frac{T_c(1)-T_c(1-1/L^2)}{1/L^2}\,,
\end{equation}
corresponding to a single missing site in an $L\times L$ supercell \cite{AuYang1976}.
By translational symmetry, all realizations are equivalent and a single configuration suffices.
The results converge monotonically to the exact analytical value:
\begin{eqnarray}
T'_c(1)|_{L=1000} &=& 3.550813\dots\nonumber\\
T'_c(1)|_{L=2000} &=& 3.550794\dots\nonumber\\
T'_c(1)|_{L=3000} &=& 3.550789\dots\nonumber\\
T'_c(1)|_{L=4000} &=& 3.550787\dots\label{Tprime}
\end{eqnarray}
The agreement with the exact result to six decimal digits demonstrates the high precision and rapid convergence of our method.
Higher derivatives and the general expansion in $q=1-p$ around the Ising point can be computed analogously \cite{AuYang1976}.

\vspace{5pt}
\section*{Weak to moderate dilutions}
For weak to moderate dilution ($p \gtrsim \tfrac{2}{3}$), the critical temperature decreases smoothly with dilution and shows negligible dependence on the supercell size $L$, with curves for successive $L$ collapsing within numerical precision.
This rapid convergence reflects strong self-averaging: the distributions of the FV eigenvalue $\lambda_c$ are sharply peaked and approximately Gaussian, with variance decreasing as $L$ increases, see, e.g. Fig.~\ref{convergence}(a).
In this regime even a single disorder realization already provides a statistically representative estimate of the ensemble average, and therefore of $T_c(p)$.
To benchmark our method, we examined representative weak dilutions $p = \tfrac{9}{10}$, $\tfrac{3}{4}$, extensively studied in previous Monte Carlo simulations \cite{Kim1995,Ballesteros1997,Hasenbusch2008}.
Disorder-averaged critical temperatures at these concentrations were determined to seven significant digits even without extrapolation, as reported in Table~\ref{precise}; a similar precision can be achieved for {\it any} $p$ in the regime of weak to moderate dilutions.
This surpasses the accuracy of large-scale Monte Carlo studies despite requiring only modest computational resources. Detailed statistics are provided in the End Matter.
\begin{figure}[t!]
    \centering
\includegraphics[width=\columnwidth]
{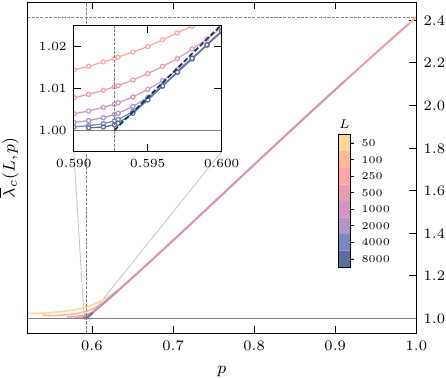}
\caption{Mean FV eigenvalue $\overline{\lambda}_c(L,p)$ of the FV matrix
as a function of the site occupation probability $p$, for increasing
supercell sizes $L$. Horizontal lines indicate the bounds
$\lambda_c(p_c) = 1$ and $\lambda_c(1) = 1+\sqrt{2}$, while the vertical
dashed line indicates the site-percolation threshold
$p_c \simeq 0.592746\ldots$. \emph{Inset:} enlargement of the
near-critical region. The dashed line shows the linear approximation
$\lambda^{\mathrm{lin}}_c(p)$ defined in Eq.~\eqref{pseudolambda}.}
\label{lambda}
\end{figure}

\renewcommand{\arraystretch}{1.5}
\begin{table}[t]
\centering\small
\label{tab:example}
\begin{tabular}{cccc}
\toprule
$p$ & $\lambda_c$ &$\beta_c$ & $T_c$ \\
\hline
$1$ & $2.414213\dots$ &$0.440686\dots$ & $\mathbf{2.269185\dots}$  \\
$\tfrac{9}{10}$ & $2.073866(1)$ &$0.5258356(3)$ & $\mathbf{1.901735(1)}$  \\
$\tfrac{3}{4}$ & $1.543883(3)$ &$0.771356(2)$ & $\mathbf{1.296417(4)}$  \\
$\tfrac{3}{5}$ & $1.023612(10)$ & $2.2259(2)$ & $\mathbf{0.44930(4)}$  \\
\bottomrule
\end{tabular}
\caption{Critical temperatures at representative values of $p$, determined to between five and seven significant digits depending on dilution. See the End Matter for detailed statistics and convergence data.}
\label{precise}
\end{table}

\vspace{5pt}
\section*{Strong dilutions}
Finite-size effects become increasingly pronounced as $p$ approaches the site percolation threshold $p_c$, where large-scale connectivity fluctuations dominate and the critical temperature drops rapidly.
In this regime, individual supercells are no longer guaranteed to contain a spanning cluster, and the resulting disorder realizations can differ qualitatively in their connectivity properties.
Each configuration is classified according to the dimensionality of its percolating cluster: two-dimensional (spanning in both directions), one-dimensional (spanning in only one direction), or zero-dimensional (non-percolating).
To accelerate sampling near $p_c$, we pre-classify configurations using a standard cluster-identification algorithm and restrict the FV calculation to those with full two-dimensional percolation, leaving the statistical outcome unchanged.
As a consistency check, at $p_c$ our sampling reproduces the universal crossing probabilities on the torus as predicted through Conformal Field Theory by Pinson \cite{Pinson1994}: the fractions $N_0/N$, $N_1/N$ and $N_2/N$ of non-percolating, singly-spanning (possibly wrapping), and doubly-spanning configurations (also known as crossing configurations)  converge respectively to the
values $\Pi_0=0.3095...$, $\Pi_1=0.3809...$ and $\Pi_2=\Pi_0$.
A depiction of crossing probabilities is provided in the End Matter.
Only the fully two-dimensional configurations yield $T_c>0$ (equivalently $\lambda_c>1$); the remaining ones have $T_c=0$ ($\lambda_c=1$).
For a faithful representation of the finite-$L$ phase boundary, these non-spanning contributions must be included in the disorder average --- they are responsible for the suppression of $\overline{T}_c(L,p)$ below $p_c$ visible in the finite-$L$ curves of Fig.~\ref{phasediagram}.
However, since the fraction of spanning configurations rises to unity for $p>p_c$ as $L$ increases, these contributions vanish in the thermodynamic limit.
Convergence is nonetheless slower near $p_c$: Fig.~\ref{convergence}(b) illustrates how at $p=\frac{3}{5}$, the eigenvalue distribution requires significantly larger supercells to stabilize.
Despite this slower convergence, our method still achieves high precision at $p = \frac{3}{5}$, as reported in Table~\ref{precise}.

\begin{figure}
    \centering
\includegraphics[width=1\columnwidth]
{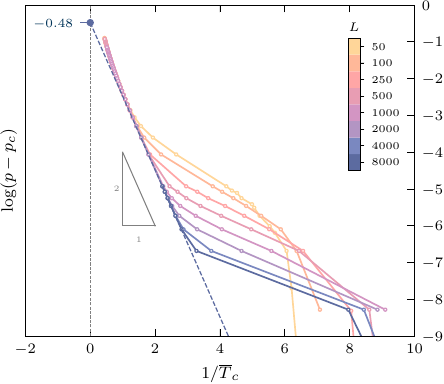}
    \caption{Scaling of the critical temperature near the percolation threshold. The plot shows $\log(p-p_c)$ versus $1/{\overline{T}}_c$ for increasing supercell sizes $L$.
    Near the percolation threshold (before the breakdown of the approximation) the data show that $T_c(p)$ approaches the expected asymptotic behavior \eqref{asymp} with $\phi_{\rm RSIM}=1$ and slope $-2$.
    The intercept from the fit (dashed line) gives the estimate $-\log \alpha_{\rm RSIM} = - 0.48$.}
    \label{scaling}
\end{figure}

\vspace{5pt}
\section*{At the percolation threshold}
From our data (inset of Fig.~\ref{lambda}) the average FV eigenvalue $\overline{\lambda}_c(L,p_c)$ approaches $\overline{\lambda}_c(L,p_c) \to 1$ as $L$ increases, implying that $\overline{T}_c(L,p_c)$ tends continuously to zero at the percolation threshold.
This confirms that the spanning cluster at $p_c$ does not sustain long--range order.

A remarkable feature of our results is that the critical eigenvalue $\lambda_c(p)$ follows to a high accuracy a straight line connecting $(p_c,1)$ to $(1,\lambda_c(1))$, where $\lambda_c(1)=\sqrt{2}+1$, see Fig.~\ref{lambda}.
Assuming exact linearity yields the following linear interpolation
\begin{equation}\label{pseudolambda}
\lambda^{\rm{lin}}_c(p) = 1 + \frac{\lambda_c(1)-1}{1-p_c}\,(p-p_c)\,,
\end{equation}
which implies the analytic approximation
\begin{equation}\label{pseudoT}
T^{\rm{lin}}_c(p) = \frac{1}{\mathrm{arctanh}\frac{1}{1+\sqrt{2}\frac{p-p_c}{1-p_c}}}\,.
\end{equation}
In the limit $p\to p_c^+$, this relation gives
\begin{equation}
T_{c}(p)=-\frac{2/\phi}{ \log [\alpha(p-p_c)]}+O(p-p_c)\,,
\label{asymp}
\end{equation}
with crossover exponent $\phi_{\rm lin}=1$ and nonuniversal amplitude $\alpha_{\rm lin} = \sqrt{2} = 1.41421...$.
The asymptotic relation \eqref{asymp} was first established for the random bond Ising model (RBIM) by Bergstresser~\cite{Bergstresser1976}, who showed that $\phi_{\rm RBIM}=1$, and sharpened by Domany~\cite{Domany1978}, who computed $\alpha_{\rm RBIM} = 2\log{2} = 1.38629...$.
The crossover exponent $\phi$ was established on general grounds by Coniglio to equal exactly one \cite{Coniglio1981}.
For the RSIM, the analysis of de Queiroz and Stinchcombe~\cite{Queiroz1992} was not sufficient to determine $\alpha_{\rm RSIM}$.
Our data (Fig.~\ref{scaling}) confirm $\phi_{\rm RSIM}=1$: plotting $\log(p-p_c)$ versus $1/\overline{T}_c$ yields a linear relation with slope $-2$,
and from the intercept we extract the nonuniversal amplitude $\alpha_{\rm RSIM} \simeq {\rm{e}}^{0.48} \approx 1.616$ providing the first reliable quantitative estimate for this quantity.

\begin{figure}[t!]
    \centering
\includegraphics[width=1\columnwidth]
{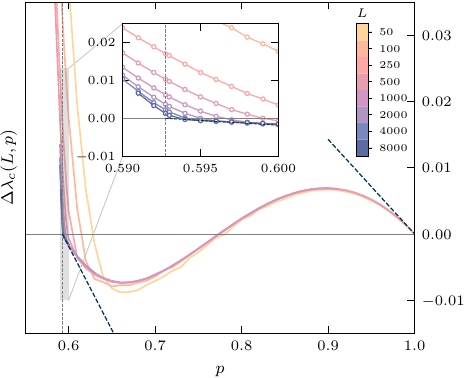}
\caption{Deviation of the mean FV eigenvalue from the linear approximation,
$\Delta\lambda_c(L,p) = \overline{\lambda}_c(L,p) - \lambda_c^{\mathrm{lin}}(p)$
for increasing supercell sizes $L$.
The vertical dashed line marks $p_c$; the horizontal line is the zero baseline.
The dashed dark lines are the analytical tangents to $\Delta\lambda_c$ at the endpoints $p = p_c$ and $p = 1$, which are directly related to $\alpha_{\rm RSIM}$ and $T'_c(1)$, respectively.
\emph{Inset:} magnification of the near-critical region $p \in [0.59,\,0.60]$.
}
\label{deltalambda}
\end{figure}
Although the linear interpolation is remarkably accurate, small systematic deviations indicate it is not exact.
For instance, it predicts $T_c^{\rm lin}(\tfrac{9}{10})=1.89416...$, $T_c^{\rm lin}(\tfrac{3}{4})=1.29908\dots$, $T_c^{\rm lin}(\tfrac{3}{5})=0.455895...$ and $T'^{\,\rm lin}_c(1)=3.70325...$, to be compared with the exact values in Tab.~\ref{precise} and Eq.~\eqref{Tprime}.
To isolate the nontrivial structure of the phase boundary, we consider the deviation from the linear interpolation $\Delta \lambda_c(p) \equiv \lambda_c(p) - \lambda_c^{\mathrm{lin}}(p)\,.
$
As shown in Fig.~\ref{deltalambda}, $\Delta \lambda_c(p)$ is small across the entire range of $p$, confirming the near-linearity of $\lambda_c(p)$, but exhibits clear systematic deviations that encode the nontrivial structure of the exact phase boundary.
To our knowledge, this fine structure is resolved here for the first time, revealing the mathematical complexity of the exact solution of the RSIM.

\vspace{5pt}
\section*{Outlook}
The approach introduced here — generalizing the combinatorial solution of the two-dimensional Ising to randomized supercells — is not restricted to site dilution on the square lattice. It applies directly to bond-diluted Ising models, where the transition matrix $\mathbb{W}$ is modified by removing bonds rather than sites, and extends naturally to any planar lattice admitting an exact FV solution, including triangular, honeycomb and Archimedean lattices \cite{Codello2010, Laurent_2025, Portillo_2025, Joseph_2026}.
Furthermore, the present method can be directly employed to compute all quenched thermodynamical quantities with similar precision, with particular interest to elucidate the behavior the specific heat at the critical point.
Beyond dilution, the technique naturally applies to the two dimensional Edwards–Anderson $\pm J$ or Gaussian spin glass -- and more broadly, the strategy of building quenched disorder into an exact lattice solution via supercells is applicable to any other integrable statistical model, opening a systematic route to attack problems in disordered systems that have so far resisted precise determination.

\vspace{5pt}
\subsection*{Acknowledgments}
Authors acknowledge fundamental compute support from the CSIC grant I+D-2022-22520220100174UD.
A.C. also acknowledges financial support from ANII-SNI-2023-1-1013433.
G.G. work is supported by the Deutsche Forschungsgemeinschaft (DFG, German Research Foundation) under Germany's Excellence Strategy EXC2181/1-390900948 (the Heidelberg STRUCTURES Excellence Cluster) and by the European Union under GA No. 101077500--QLRNet.

\contacts{%
  \safeemail{riccardo.ben.ali.zinati}{usi.ch} \\
  \safeemail{gori}{thphys.uni-heidelberg.de} \\
  \safeemail{alessandro.codello}{unive.it}%
}

\spcolophon
\bibliography{bib}

\clearpage
\onecolumn
%%%%%%%%%%%%%%%%%%
\section*{
End Matter
%Supplementary Material
}
\begin{figure}[t!]
\includegraphics[width=1\columnwidth]
{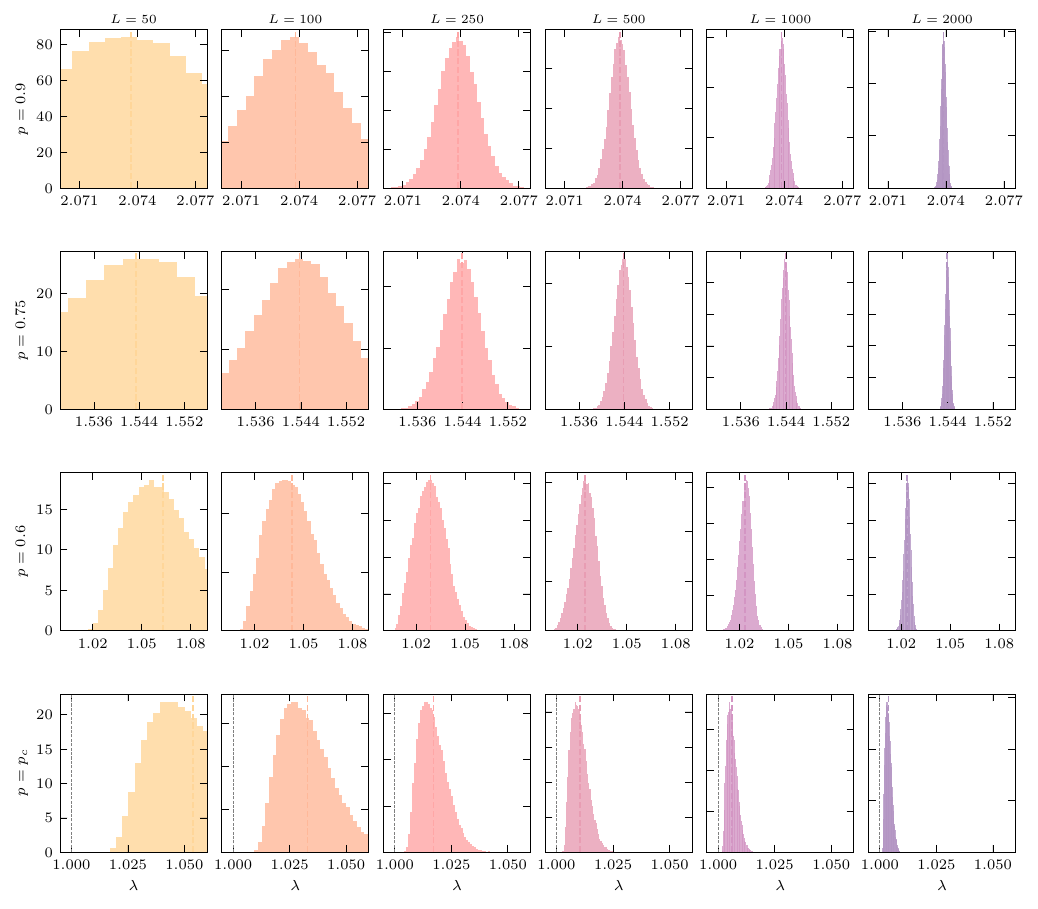}
    \caption{Probability density $\rho(\lambda_c)$ of the FV eigenvalue for different system sizes $L$ (columns) and dilution probabilities $p$ (rows).
    Histograms show the distribution over disorder realizations, while dashed colored lines indicate the sample mean $\overline{\lambda}_c$. The gray vertical line marks $\lambda_c=1$, corresponding to $T_c=0$.}
    \label{table_convergence}
\end{figure}

\noindent
Here we report the numerical data used in the main analysis for the representative
dilutions $p=0.9$, $0.75$, $0.6$, and $p_c$.
All computations were performed in base-10 to avoid rounding artifacts.
For each supercell size $L$ the tables list
the number of non-spanning ($N_0$), singly-spanning ($N_1$), and doubly-spanning ($N_2$)
configurations obtained in the disorder sampling. 
We also report the leading FV eigenvalue averaged over the percolating
($\mathrm{dim}=2$) realizations, $\overline{\lambda}_c|_2$, together with its
standard deviation $\sigma_{\lambda_c}|_2$, skewness $\zeta_{\lambda_c}|_2$, and excess
kurtosis $(\kappa_{\lambda_c}-3)|_2$.
In addition, the tables provide the disorder-averaged eigenvalue $\overline{\lambda}_c$, the corresponding inverse temperature
$\overline{\beta}_c$, and the critical temperature $\overline{T}_c$.
The corresponding distributions are shown in Fig.~\ref{table_convergence}.

\vspace{2pt}
\setlength{\floatsep}{4pt plus 1pt minus 1pt}
\setlength{\textfloatsep}{4pt plus 1pt minus 1pt}
%p=0.9
\renewcommand{\arraystretch}{1.4}
\begin{table}[!h]
\centering
\resizebox{\textwidth}{!}{
\begin{tabular}{|c|c|c|c|c|c|c|c|c||c|c|c|}
\hline
$L$ & $N_0$ & $N_1$ & $N_2$ & $N_2/N$ &$\overline{\lambda}_c|_2$ & $\sigma_{\lambda_c}|_2$ & $\zeta_{\lambda_c}|_2$ & $(\kappa_{\lambda_c}-3)|_2$ & $\overline{\lambda}_c$ & $\overline{\beta}_c$ & $\overline{T}_c$\\
\hline
50 & 0 & 0 & 100\,040 &1& 2.07366742 & 0.004725 &  0.104635 & 0.087482 & 2.07367(1) & 0.525900(5) & {\bf 1.90152(2)}\\
100 & 0 & 0 & 100\,020 &1& 2.07381416 & 0.002471 & 0.058554 & 0.078973 & 2.073814(8) & 0.525852(2) & {\bf 1.901679(9)} \\
250 & 0 & 0 & 100\,020 &1& 2.07385449 & 0.001028 & 0.026790 & 0.037633 & 2.073854(3) & 0.5258391(10) & {\bf 1.901723(4)} \\
500 & 0 & 0 & 100\,020 &1& 2.07386035 & 0.000528 & 0.008768 & 0.000788 & 2.073860(2) & 0.5258372(5) & {\bf 1.901730(2)}\\
1000 & 0 & 0 & 50\,016 &1& 2.07386435 & 0.000269 & -0.001935 & 0.008962 & 2.073864(1) & 0.5258359(4) & {\bf 1.901734(1)}\\
2000 &  0 & 0 & 25\,023 &1& 2.07386550 & 0.000136 & 0.005128 & -0.072753 & 2.073866(1) & 0.5258356(3) & {\bf 1.901735(1)}\\
% \hline
% $\infty$ & -- & -- & -- & --& -- & -- & -- & -- & -- & -- & -- \\
\hline
\end{tabular}}
\caption{Data for $p=0.9$.}
\label{tablep9over10}
\end{table}

%p=0.75
\renewcommand{\arraystretch}{1.4}
\begin{table}[!h]
\centering
\resizebox{\textwidth}{!}{
\begin{tabular}{|c|c|c|c|c|c|c|c|c||c|c|c|}
\hline
$L$ & $N_0$ & $N_1$ & $N_2$ & $N_2/N$ & $\overline{\lambda}_c|_2$ & $\sigma_{\lambda_c}|_2$ & $\zeta_{\lambda_c}|_2$ & $(\kappa_{\lambda_c}-3)|_2$ & $\overline{\lambda}_c$ & $\overline{\beta}_c$ & $\overline{T}_c$\\
\hline
50 & 0 & 0 & 100\,000 & 1& 1.54341228 & 0.015461 & -0.187462 & 0.220542 &1.54341(5)&  0.77189(4) & {\bf 1.29579(6)}\\
100 & 0 & 0 & 100\,000 &1& 1.54376840 & 0.007980 & -0.086179 & 0.075156 & 1.54377(3) &  0.77149(2) & {\bf 1.29626(3)}\\
250 & 0 & 0 & 100\,000 &1& 1.54386154 & 0.003279 & -0.035256 & 0.019767 & 1.54386(1) & 0.771381(7) & {\bf  1.29639(1)}\\
500 & 0 & 0 & 100\,000 &1& 1.54387480 & 0.001666 & -0.011060 & 0.020471 &1.543875(5)& 0.771365(4) & {\bf 1.296407(6)}\\
1000 & 0 & 0 & 50\,000 &1& 1.54387867 & 0.000845 & 0.000792 & -0.008578 &1.543879(4)& 0.771360(3) & {\bf 1.296412(5)}\\
2000 & 0 & 0 & 25\,020 &1& 1.54388289 & 0.000425 & 0.019087 & -0.056880 &1.543883(3)&  0.771357(3) & {\bf 1.296417(4)}\\
4000 & 0 & 0 & 5\,000 &1& 1.54388292 & 0.000218 & 0.017839 & -0.008383 &1.543883(3)&  0.771356(2) & {\bf 1.296417(4)}\\
% \hline
% $\infty$ & -- & -- & -- & --&& &&& -- &&\\
\hline
\end{tabular}}
\caption{Data for $p=0.75$.}
\label{tablep75over100}
\end{table}

%p=0.6
\renewcommand{\arraystretch}{1.4}
\begin{table}[!h]
\centering
\resizebox{\textwidth}{!}{
\begin{tabular}{|c|c|c|c|c|c|c|c|c||c|c|c|}
\hline
$L$ & $N_0$ & $N_1$ & $N_2$ & $N_2/N$ & $\overline{\lambda}_c|_2$ & $\sigma_{\lambda_c}|_2$ &
$\zeta_{\lambda_c}|_2$ & 
$(\kappa_{\lambda_c}-3)|_2$ &  
$\overline{\lambda}_c$ & $\overline{\beta}_c$ & $\overline{T}_c$\\
\hline
50 & 42\,490 & 94\,594 & 100\,008 &0.422& 1.06286925 & 0.021070 & 0.510042 & -0.052926 &1.02652(7)&  0.748(2) & {\bf 0.2399(6)}\\
100 & 24\,778 & 67\,239 &100\,008 &0.521& 1.04289654 & 0.014446 &  0.427725 & -0.169388 &1.02234(5)&  1.022(2) & {\bf 0.2676(6)}\\
250 & 6\,574 & 31\,877 & 100\,008 &0.722& 1.02884238 & 0.009130 & 0.202380 & -0.359390 &1.02083(4)&  1.556(3) & {\bf  0.3373(6)}\\
500 & 843 & 9\,934 & 100\,008 &0.902& 1.02442381 & 0.006396 &  -0.106869 & -0.263153 &1.02205(3)&  2.012(2) &{\bf 0.4067(4)} \\
1000 & 7 & 636 & 75\,024 &0.991& 1.02350287 & 0.003894 & -0.364688 & 0.258197 &1.02330(2)&  2.2163(8) & {\bf 0.4442(2)}\\
2000 & 0 & 2 & 50\,064 &0.999& 1.02356365 & 0.002005 & -0.263986 & 0.228538 &1.023563(9)&   2.2282(2) & {\bf 0.44892(4)}\\
4000 & 0 & 0 & 10032 &1& 1.02361214 & 0.000997 & -0.123208 & 0.099277 &1.023612(10)&  2.2259(2) &  {\bf 0.44930(4)}\\
% \hline
% $\infty$ & -- & -- & -- & --&& & -- &&&&\\
\hline
\end{tabular}}
\caption{Data for $p=0.6$.}
\label{tablep6over10}
\end{table}
%
%p=pc
\renewcommand{\arraystretch}{1.4}
\begin{table}[!h]
\centering
\resizebox{\textwidth}{!}{
\begin{tabular}{|c|c|c|c|c|c|c|c|c||c|c|c|}
\hline
$L$ & $N_0$& $N_1$ & $N_2$ & $N_2/N$ & $\overline{\lambda}_c|_2$ & $\sigma_{\lambda_c}|_2$ &
$\zeta_{\lambda_c}|_2$ &
$(\kappa_{\lambda_c}-3)|_2$ &
$\overline{\lambda}_c$&
$\overline{\beta}_c$ & $\overline{T}_c$\\
\hline
50 & 104\,868 & 149\,500 & 100\,020 &0.282& 1.05369780  &  0.018310 & 0.660780 & 0.226400 &1.01516(4)&  0.522(1) & {\bf 0.1537(4)}\\
100 & 109\,315 & 143\,652 & 100\,020 &0.283& 1.03289940 & 0.011558 & 0.664070 & 0.195513 &1.00932(3)&  0.593(2) & {\bf 0.1363(4)}\\
250 & 111\,315 & 139\,078 & 100\,020 &0.285& 1.01707317 & 0.006138 & 0.671228 & 0.227408 &1.00487(1)&  0.690(2) & {\bf 0.1187(3)}\\
500 & 113\,020 & 137\,457 & 100\,020 &0.285& 1.01033476 & 0.003781 & 0.688571 & 0.302149 &1.002949(9)&  0.762(2) & {\bf 0.1074(3)}\\
{1000} & 57589 & 68585 & 50\,040 &0.283& 1.00619921 & 0.002290 & 0.710221 & 0.321996 &1.001760(7)&  0.830(3) & {\bf 0.0975(4)}\\
2000 & 29\,228 & 34\,350 & 25\,100 &0.283& 1.00373524 & 0.001370 & 0.640414 & 0.128349 &1.001057(6)&  0.899(5) & {\bf 0.0894(5)}\\
4000 & 11\,982 & 13\,585 & 10\,040 &0.282&  1.00224880 & 0.000849 & 0.661651 & 0.205307 &1.000634(6)& 0.968(8) &{\bf 0.0824(7)}\\
% \hline
% $\infty$ & -- & -- & -- & -- & {\bf 1.00000}&& & -- &&&{\bf 0.00000}\\
\hline
\end{tabular}}
\caption{Data for $p=p_c=0.592746\dots$.}
\label{tablep2over3}
\end{table}

\renewcommand{\arraystretch}{1.1}
\begin{table}[!h]
\caption{
Disorder-averaged eigenvalues $\overline{\lambda}_c$ at each supercell size $L$, with $N_2 = 10\,000$ spanning realizations per entry. Bold entries indicate values that have converged to five significant digits; the remaining entries near $p_c$ have not yet reached full convergence at the accessible supercell sizes. An asterisk ($*$) indicates that sampling was stopped early because the eigenvalue had converged to $1$ within numerical precision (non-percolating regime). At each concentration the standard error of the mean is below the last displayed digit; full error analysis at representative values of $p$ is provided in Tables~\ref{tablep9over10}--\ref{tablep6over10}.}
\centering
\scalebox{0.8}{
\begin{tabular}{|c|c|c|c|c|c|c|c|c|c|c|}
\hline
$p$ & $\overline{\lambda}_{50}$ &
$\overline{\lambda}_{100}$ &
$\overline{\lambda}_{250}$ &
$\overline{\lambda}_{500}$ &
$\overline{\lambda}_{1000}$ &
$\overline{\lambda}_{2000}$ &
$\overline{\lambda}_{4000}$ &
$\overline{\lambda}_{8000}$ \\
\hline
0.500 &1.01712*&{\bf 1.00000*}&---&---&---&---&---&---\\
0.510 &1.02250*&{\bf 1.00000*}&---&---&---&---&---&---\\
0.520 &1.02332*&{\bf 1.00000*}&---&---&---&---&---&---\\
0.530 &1.02264*&{\bf 1.00000*}&---&---&---&---&---&---\\
0.540 &1.02487*&1.01512*&{\bf 1.00000*}&---&---&---&---&---\\
0.550 &1.02770*&1.01306*&{\bf 1.00000*}&---&---&---&---&---\\
0.560 &1.03108&1.01467*&{\bf 1.00000*}&---&---&---&---&---\\
0.570 &1.03561&1.01760*&1.00728*&{\bf 1.00000*}&---&---&---&---\\
0.580 &1.04138&1.02216&1.00867*&1.00400*&{\bf 1.00000*}&---&---&---\\
0.590 &1.05067&1.02998&1.01438&1.00774&1.00392&1.00186*&1.00080*&---\\
0.591 &1.05185&1.03073&1.01523&1.00854&1.00458&1.00229&1.00107*&---\\
0.592 &1.05292&1.03204&1.01631&1.00951&1.00545&1.00299&1.00156&1.00077\\
0.592746 &1.05370&1.03290&1.01707&1.01033&1.00620&1.00373&1.00225&1.00130\\
0.593 &1.05374&1.03315&1.01743&1.01061&1.00655&1.00404&1.00255&1.00165\\
0.594 &1.05513&1.03452&1.01866&1.01197& 1.00793&1.00567&1.00446&{\bf 1.004}06\\
0.595 &1.05616&1.03560&1.01994&1.01349&1.00973&1.00785&1.00731&{\bf 1.007}22\\
0.596 &1.05755&1.03710&1.02157&1.01521&1.01182&1.01063&1.01050&{\bf 1.010}56\\
0.597 &1.05878&1.03870&1.02318&1.01696&1.01439&1.01375&1.01384&{\bf 1.013}82\\
0.598 &1.06021&1.03968&1.02481&1.01933& 1.01722&1.01698&1.01706&{\bf 1.017}13\\
0.599 &1.06201&1.04141&1.02635&1.02177&1.02025&1.02027&1.02032&{\bf 1.020}37\\
0.600 &1.06287&1.04290&1.02884&1.02442&1.02350&1.02356&1.02361&{\bf 1.023}68\\
0.610 &1.07929&1.06367&1.05631&1.05641&1.05647&{\bf 1.0565}4&---&---\\
0.620 &1.10130&1.09054&1.08944&1.08971&1.08986&{\bf 1.0898}5&---&---\\
0.630 &1.12789&1.12257&1.12320&1.12343&1.12349&{\bf 1.1234}9&---&---\\
0.640 &1.15764&1.15682&1.15734&1.15737&1.15747&{\bf 1.1575}2&---&---\\
0.650 &1.19064&1.19090&1.19159&1.19163&{\bf 1.1917}4&---&---&---\\
0.660 &1.22473&1.22580&1.22616&1.22615&{\bf 1.2262}2&---&---&---\\
0.670 &1.25951&1.26054&1.26094&1.26095&{\bf 1.2609}8&---&---&---\\
0.680 &1.29452&1.29564&1.29586&1.29591&{\bf 1.2959}2&---&---&---\\
0.690 &1.33000&1.33066&1.33094&1.33100&{\bf 1.3310}1&---&---&---\\
0.700 &1.36536&1.36602&1.36636&1.36628&{\bf 1.36628}&---&---&---\\
0.710 &1.40075&1.40167&1.40164&1.40163&{\bf 1.40164}&---&---&---\\
0.720 &1.43635&1.43689&1.43706&1.43713&{\bf 1.43712}&---&---&---\\
0.730 &1.47162&1.47248&1.47260&1.47264&{\bf 1.47266}&---&---&---\\
0.740 &1.50775&1.50820&1.50821&1.50822&{\bf 1.50825}&---&---&---\\
0.750 &1.54341&1.54376&1.54386&1.54387&1.54388&1.54388&{\bf 1.54388}&---\\
0.760 &1.57910&1.57947&1.57947&1.57950&{\bf 1.57952}&---&---&---\\
0.770 &1.61490&1.61511&1.61520&1.61518&{\bf 1.61518}&---&---&---\\
0.780 &1.65016&1.65071&1.65090&1.65082&{\bf 1.65082}&---&---&---\\
0.790 &1.68602&1.68638&1.68641&1.68644&{\bf 1.68644}&---&---&---\\
0.800 &1.72184&1.72199&1.72202&1.72203&{\bf 1.72202}&---&---&---\\
0.810 &1.75715&1.75734&1.75755&1.75757&{\bf 1.75754}&---&---&---\\
0.820 &1.79285&1.79291&1.79299&1.79304&{\bf 1.79303}&---&---&---\\
0.830 &1.82807&1.82834&1.82843&1.82845&{\bf 1.82844}&---&---&---\\
0.840 &1.86347&1.86379&1.86378&1.86377&{\bf 1.86378}&---&---&---\\
0.850 &1.89864&1.89896&1.89901&1.89902&{\bf 1.89902}&---&---&---\\
0.860 &1.93390&1.93414&1.93419&1.93420&{\bf 1.93420}&---&---&---\\
0.870 &1.96895&1.96923&1.96925&1.96927&{\bf 1.96927}&---&---&---\\
0.880 &2.00400&2.0042&2.00423&2.00424&{\bf 2.00424}&---&---&---\\
0.890 &2.03885&2.03904&2.03910&2.03911& {\bf 2.03911}&---&---&---\\
0.900 &2.07366&2.07381&2.07385&2.07386&2.07386&{\bf 2.07386}&---&---\\
0.910 &2.10832&2.10843&2.10847&2.10850&{\bf 2.10850}&---&---&---\\
0.920 &2.14278&2.14300&2.14300&2.14301&{\bf 2.14301}&---&---&---\\
0.930 &2.17725&2.17738&2.17739&{\bf 2.17740}&---&---&---&---\\
0.940 &2.21147&2.21161&2.21165&{\bf 2.21166}&---&---&---&---\\
0.950 &2.24562&2.24575&2.24577&{\bf 2.24578}&---&---&---&---\\
0.960 &2.27966&2.27973&2.27977&{\bf 2.27977}&---&---&---&---\\
0.970 &2.31350&2.31357&2.31360&{\bf 2.31360}&---&---&---&---\\
0.980 &2.34721&2.34727&2.34729&{\bf 2.34730}&---&---&---&---\\
0.990 &2.38079&2.38082&2.38083&{\bf 2.38083}&---&---&---&---\\
1.000 &2.41421&2.41421&2.41421&{\bf 2.41421}&---&---&---&---\\
\hline
\end{tabular}
}
\label{table_eigs}
\end{table}

\begin{figure}[h!]
    \centering
\includegraphics[width=\textwidth]
{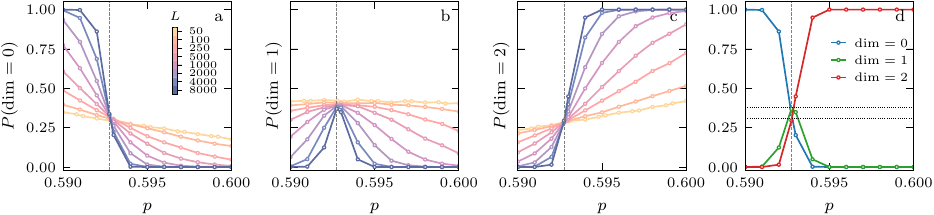}
\caption{Probabilities of different spanning classes as a function of the site occupation probability $p$ for increasing supercell sizes $L$. Panels (a–c) show the disorder-averaged probabilities of non-spanning ($\mathrm{dim}=0$), singly-spanning ($\mathrm{dim}=1$), and doubly-spanning ($\mathrm{dim}=2$) configurations, respectively. Colors indicate the supercell size $L$. Panel (d) shows the three probabilities for the largest system size, illustrating their crossing near the percolation threshold $p_c$. Horizontal dotted lines indicate the universal Pinson crossing probabilities $\Pi_0 = \Pi_2\simeq 0.3095\dots$ and $\Pi_1 \simeq 0.3809\dots$, while the vertical dashed line marks $p_c$.}
\label{counts}
\end{figure}

\begin{figure}[h]
    \centering
\includegraphics[width=\textwidth]
{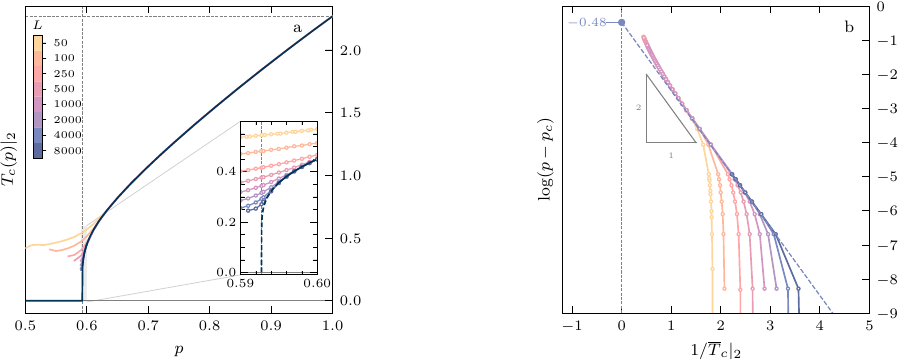}
\caption{(a):
Phase boundary $T_c(p)|_2$ of the randomly site-diluted Ising model computed for increasing disordered supercell sizes 
$L$, conditioned on doubly-spanning clusters configurations.
The inset magnifies the critical region near the percolation threshold $p_c$.
The unrestricted version given in Fig.~\ref{phasediagram} corresponds to $\overline{T}_c=\frac{N_2}{N}\overline{T}_c|_2$.
(b):  Scaling of the restricted critical temperature near the percolation threshold. The plot shows $\log(p-p_c)$ versus $1/{\overline{T}}_c$ for increasing supercell sizes $L$.
Near the percolation threshold (before the breakdown of the approximation) the data show that $T_c(p)$ approaches the expected asymptotic behavior \eqref{asymp} with $\phi_{\rm RSIM} =1$ and slope $-2$.
The intercept from the fit (dashed line) gives the estimate $\log \alpha_{\rm RSIM} = 0.48$.}
\label{counts}
\end{figure}

\end{document}